\documentclass{ws-ijmpa}

\begin{document}

\markboth{Lothar Tiator} {Photoproduction of eta and etaprime mesons}

\title{PHOTOPRODUCTION OF ETA AND ETAPRIME MESONS ON THE NUCLEON}

\author{LOTHAR TIATOR}

\address{Institut f\"ur Kernphysik, Universit\"at Mainz\\
D-55099 Mainz, Germany\\
tiator@kph.uni-mainz.de}

\maketitle

\begin{abstract}
The isobar models $\eta$-MAID and $\eta'$-MAID have been used to analyze new data on
quasi-free $\eta$ photoproduction on the deuteron from Bonn and recent $\eta'$ data on
the proton from Jlab. In $\eta$ photoproduction on the neutron a bump around $W=1700$~MeV
was observed which could possibly arise from a narrow $P_{11}$ state that is discussed as
a non-strange member of the $\Theta^+$ antidecuplett. In $\eta'$ photoproduction on the
proton resonance contributions are found that can be attributed to missing resonances in
the energy region around $W=1900$~MeV.

\keywords{Eta Meson; Photoproduction; Nucleon Resonances.}
\end{abstract}

\ccode{PACS numbers: 13.60.Le, 14.20.Gk, 25.20.Lj, 25.30.Rw}

\section{Introduction}

In recent experiments of eta photoproduction on the deuteron at
GRAAL\cite{Kouznetsov:2004} and at CB-TAPS-ELSA\cite{Jaegle:2006} an enhanced cross
section has been observed in the neutron channel around $W=1700$~MeV, which is not
visible in eta photoproduction on the proton target. In 1997, Diakonov, Petrov and
Polyakov had predicted an exotic anti-decuplet of baryons within the chiral soliton
model\cite{Diakonov:1997}. Besides the famous $\Theta^+(1530)$, they also predicted a
non-strange member $N(1710)$, which could have been identified with the $P_{11}(1710)$
state listed in the particle data tables\cite{PDG2004}. Furthermore, this state was
predicted with a much stronger coupling to the $\eta N$ than to the $\pi N$ channel.
After the first pentaquark signals had been reported\cite{Nakano:2003}, in a new partial
wave analysis Arndt et al.\cite{Arndt:2004} reported about two possible non-strange
candidates with masses 1680~MeV and 1730~MeV and widths smaller than 30~MeV. Before,
Polyakov and Rathke\cite{Polyakov:2003} had shown that within their model, the
non-strange member of the antidecuplett around 1700~MeV should couple much stronger
electromagnetically to the neutron than to the proton. In this respect the new
observations received some attractions and explanations with pentaquark states have been
searched. However, very different to the $\Theta^+$ signals that were reported as
extremely narrow peaks of the order of and even below 1~MeV, the bump in eta
photoproduction is quite broad and looks more like a typical nucleon resonance structure
with a width of the order of 100~MeV.

Surprisingly, the isobar model EtaMaid2001\cite{Chiang:2002} which was only fitted to
proton data available in 2001, describes this bump structure very well. The reason is
that the $D_{15}(1675)$ resonance came out of this fit with an unusually large $\eta N$
branching ratio, strongly violating $SU(3)$ symmetry bounds\cite{Polyakov:2006}. Already
from pion photoproduction and also from the simple quark model, the $D_{15}$ resonance is
well known as a resonance that couples much stronger to the neutron than to the proton.
As an alternative to the strong $D_{15}$ model, in our more recent isobar analysis
(EtaMaid2003) we have found that a modification of the nonresonant background, however,
would modify the eta branching very strongly. In Tables~\ref{ta1} and \ref{ta2} it is
shown that some higher resonances (and in particular the $D_{15}$) almost disappear if
the standard treatment of $t$-channel vector meson exchange with single poles is replaced
by Regge trajectories. Such a reggeization is definitely required at very large energies
of a few GeV, but at the rather low energies of $E_\gamma\approx 1$~GeV, where the bump
occurs, Regge trajectories are usually not seriously considered.

\section{Isobar models for $\eta$ and $\eta'$ photoproduction}

The isobar models $\eta$-MAID and $\eta'$-MAID are similar to the unitary isobar model
MAID. They are constructed with a nonresonant background of nucleon Born terms and
$t$-channel vector meson exchange, plus a number of $s$-channel nucleon resonance
excitations,
\begin{equation}
t_{\gamma,\eta}^\alpha = v_{\gamma,\eta}^\alpha (Born +
\omega,\rho) +t_{\gamma,\eta}^\alpha(Resonances)\,.
\end{equation}
The nonresonant background contains the usual Born terms and vector meson exchange
contributions. It is obtained by evaluating the Feynman diagrams derived from an
effective Lagrangian. The Born terms are evaluated in the standard way with pseudoscalar
coupling, and the details can be found in Ref.~\refcite{Chiang:2002}. In the reggeized
model, however, we do not include the Born terms. The reason is that the correct
treatment for the $u$-channel nucleon exchange, together with the reggeized $t$-channel
vector meson exchanges, requires to also introduce the nucleon Regge trajectories.
Because of the lack of high energy data at backward angles, it is currently difficult to
determine this $u$-channel contribution. Since the coupling constants $g_{\eta NN}$ and
$g_{\eta' NN}$ are small, the difference caused by the absence of the Born terms is
negligible at low energies.

For each partial wave $\alpha$ the resonance excitation is parameterized with standard
Breit-Wigner functions with energy dependent widths,
\begin{equation} \label{eq:BWres}
 t_{\gamma,\eta}^\alpha(R\,; \lambda) = \tilde{A}_{\lambda}\,
 \frac{\Gamma_{tot}\,W_R}{W_R^2-W^2-iW_R\Gamma_{tot}}\,
 f_{\eta N}(W)\,C_{\eta N}\; \zeta_{\eta N}\,,
\end{equation}
where a hadronic phase is introduced, $\zeta_{\eta N}=\pm1$, a relative sign between the
$N^* \rightarrow \eta N$ and the $N^* \rightarrow \pi N$ couplings. For a few states the
relative phases are well determined and can be found in the Particle Data Tables, for
most of the states it can be used as a free parameter in our partial wave (pw) analysis.
The principal fit parameters of our pw analysis are the resonance masses $W_R\equiv M^*$,
the total widths $\Gamma_R=\Gamma_{tot}(W_R)$, the branching ratios $\beta_{\eta
N}=\Gamma_{\eta N}(W_R)/\Gamma_R$ and the photon couplings
$\tilde{A}_{\lambda}=\{A_{1/2},A_{3/2}\}$. However, we fix those parameters, where
reliable results are given by PDG, see Tables~\ref{ta1} and~\ref{ta2}.

The total width $\Gamma_\mathrm{tot}$ in Eq.~(\ref{eq:BWres}) is the sum of $\Gamma_{\eta
N}$, the single-pion decay width $\Gamma_{\pi N}$, and the rest, for which we assume
dominance of the two-pion decay channels,
\begin{equation}
 \Gamma_\mathrm{tot}(W) =
 \Gamma_{\eta N}(W) + \Gamma_{\pi N}(W) + \Gamma_{\pi\pi N}(W)\,.
\end{equation}
The details of the parametrization of the energy-dependent widths and the vertex function
$f_{\eta N}(W)$ can be found in Refs.~\refcite{Chiang:2002,Chiang:2003}.

\section{Results}

\subsection{$\bm{\eta}$ photoproduction results on protons and neutrons}
\label{sec:eta}%
In this section, we present the $\eta$ photoproduction results from the reggeized model
as well as the standard $\eta$-MAID model with vector meson pole contributions. In the
reggeized model, we replace the $t$-channel $\rho$ and $\omega$ exchanges used in
$\eta$-MAID by the Regge trajectories while keeping the same $N^*$ contributions. Both
models are fitted to photoproduction data of cross sections from TAPS\cite{Krusche:1995},
GRAAL\cite{Renard:2000}, and CLAS\cite{Dugger:2002} as well as polarized beam asymmetries
from GRAAL\cite{Ajaka:1998}.

\begin{table}[htb]
\tbl{Parameters of nucleon resonances from EtaMaid2003 with standard vector meson poles,
model (I). The masses and widths are given in MeV, $\beta_{\eta N}$ is the branching
ratio for the eta decay channel and $\zeta_{\eta N}$ the relative sign between the $N^*
\rightarrow \eta N$ and the $N^* \rightarrow \pi N$ couplings. The photon couplings to
the proton and neutron target for helicity $\lambda=$1/2 and 3/2 are given in units of
$10^{-3}/\sqrt{GeV}$. The underlined parameters are fixed and are taken from
PDG2004\protect\cite{PDG2004}. The asterisk for $_nA_{1/2}$ of the $S_{11}(1535)$ denotes
a fixed $n/p$ ratio obtained from the experiment\protect\cite{Krusche:1995deu}.}
{\begin{tabular}{ccccccccc} \toprule
 $N^*$ & Mass & Width & $\beta_{\eta N}$ & $\zeta_{\eta N}$ & $_pA_{1/2}$ & $_pA_{3/2}$ & $_nA_{1/2}$ & $_nA_{3/2}$ \\
\colrule
 $D_{13}(1520)$ & \underline{1520}  & \underline{120} & $0.05\%$ & $+1$ &  -39 & \underline{166} & \underline{-59} & \underline{-139} \\
 $S_{11}(1535)$ & 1545  & 203 & $  \underline{50\%}$ & $+1$ & 125 &  -  & -102$^*$  &   -  \\
 $S_{11}(1650)$ & 1640  & 130 & $  10\%$ & $-1$ &  {73} &  -  & {-59} &   -  \\
 $D_{15}(1675)$ & 1682  & \underline{150} & $ 17 \%$ & $-1$ &  17 &  24 & \underline{-43} &  \underline{-58} \\
 $F_{15}(1680)$ & 1670  & \underline{130} & $0.04 \%$ & $+1$ & -9  & 145 &  \underline{29} &  \underline{-33} \\
 $D_{13}(1700)$ & \underline{1700}  & \underline{100} & $0.7  \%$ & $-1$ & \underline{-18} &  \underline{-2} &  \underline{0}  &   \underline{-3} \\
 $P_{11}(1710)$ & 1725  & \underline{100} & $  26\%$ & $+1$ &  22 &  -  &  \underline{-2} &   -  \\
 $P_{13}(1720)$ & \underline{1720}  & \underline{150} & $ 6.6\%$ & $-1$ &  \underline{18} & \underline{-19} &   \underline{1} &  \underline{-29} \\
\botrule
\end{tabular} \label{ta1}}
\end{table}
%%%%%%%%%%%%%%%%%%%%%%%%%%%%%%%% end table %%%%%%%%%%%%%%%%%%%%%%%%%%%%%%%%%
%%%%%%%%%%%%%%%%%%%%%%%%%%%%%%%%%% table %%%%%%%%%%%%%%%%%%%%%%%%%%%%%%%%%%%

\begin{table}[htb]
\tbl{Parameters of nucleon resonances from EtaMaid2003 with reggeized vector mesons,
model (II). Notation as in Table~\protect\ref{ta1}.} {\begin{tabular}{ccccccccc} \toprule
 $N^*$ & Mass & Width & $\beta_{\eta N}$ & $\zeta_{\eta N}$ & $_pA_{1/2}$ & $_pA_{3/2}$ & $_nA_{1/2}$ & $_nA_{3/2}$ \\
\colrule
 $D_{13}(1520)$ & \underline{1520}  & \underline{120} & $0.04\%$ & $+1$ & \underline{-24} & \underline{166} & \underline{-59} & \underline{-139} \\
 $S_{11}(1535)$ & 1521  & 118 & $  \underline{50\%}$ & $+1$ &  80 &  -  &  -65$^*$  &   -  \\
 $S_{11}(1650)$ & 1635  & 120 & $  16\%$ & $-1$ &  \underline{46} &  -  & \underline{-38} &   -  \\
 $D_{15}(1675)$ & 1665  & \underline{150} & $ 0.7\%$ & $+1$ &  19 &  15 & \underline{-43} &  \underline{-58} \\
 $F_{15}(1680)$ & 1670  & \underline{130} & $0.003\%$ & $+1$ & -15 & 133 &  \underline{29} &  \underline{-33} \\
 $D_{13}(1700)$ & \underline{1700}  & \underline{100} & $0.025\%$ & $-1$ & \underline{-18} &  \underline{-2} &  \underline{0}  &   \underline{-3} \\
 $P_{11}(1710)$ & 1700  & \underline{100} & $  26\%$ & $-1$ & \underline{9} &  -  &  \underline{-2} &   -  \\
 $P_{13}(1720)$ & \underline{1720}  & \underline{150} & $   4\%$ & $+1$ &  \underline{18} & \underline{-19} &   \underline{1} &  \underline{-29} \\
\botrule
\end{tabular} \label{ta2}}
\end{table}
%%%%%%%%%%%%%%%%%%%%%%%%%%%%%%%% end table %%%%%%%%%%%%%%%%%%%%%%%%%%%%%%%%%

\begin{figure}[htp]
\centerline{
\includegraphics[width=6.2cm, angle=0]{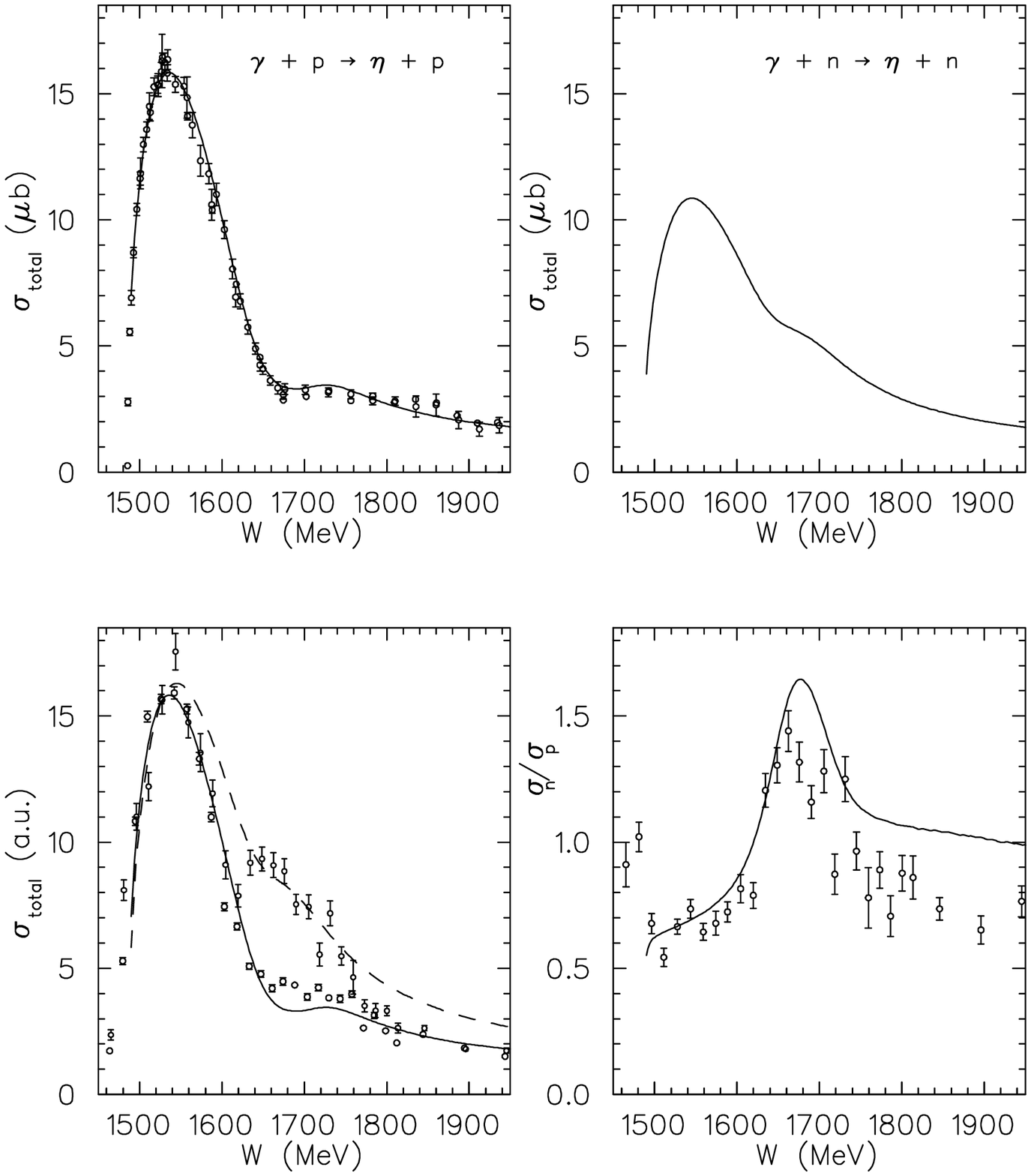}
\includegraphics[width=6.2cm, angle=0]{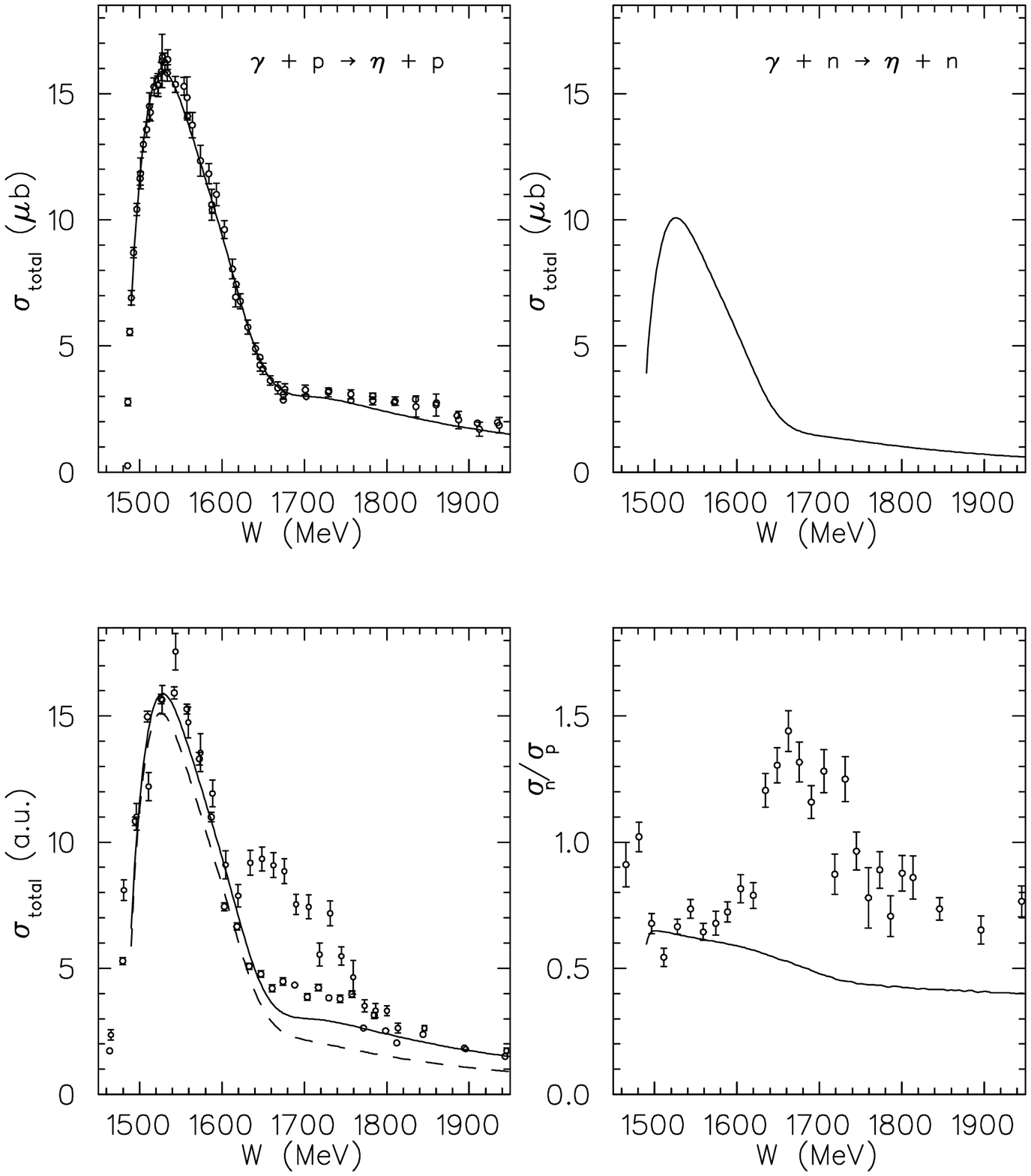}}
\caption{\label{fig:CBeta} Total cross sections for eta
photoproduction on proton and neutron (left: Maid2003 with standard vector meson poles,
right: Maid2003 with reggeized vector mesons). In the upper row the calculations are
compared to published data on the proton target from MAMI\protect\cite{Krusche:1995} (in
larger energy bins), GRAAL\protect\cite{Renard:2002} (without the five largest energy
points), CLAS\protect\cite{Dugger:2002} and CB-ELSA\protect\cite{Crede:2005}. In the
lower row the calculations are compared to preliminary data of
CB-TAPS-ELSA\protect\cite{Jaegle:2006}, shown in arbitrary units. }
\end{figure}

In Fig.~\ref{fig:CBeta} we compare our results with published data (upper panels) and
with the preliminary results of the CB-TAPS-ELSA experiment (lower panels) for both
EtaMaid2003 versions. The model using standard vector meson poles, similar to the
EtaMaid2001 model, produces a bump in the neutron cross section, whereas the model with
vector meson Regge trajectories shows similar structures for proton and neutron cross
sections and leads therefore to a flat neutron to proton ratio.

\begin{table}[ht]
\tbl{Mass, total width, $\eta N$ branching ratio and photon helicity couplings in units
of $(10^{-3}/\sqrt{GeV})$ for the $P_{11}$ pentaquark state in our calculation.}
{\begin{tabular}{ccccc}
\toprule
$M^*$ (MeV) & $\Gamma_{tot}$ (MeV) & $\Gamma_{\eta N}/\Gamma_{tot}$ & $A_{1/2}(p)$ & $A_{1/2}(n)$ \\
\colrule
 1675 & 10 & 40\% & 10 & 30 \\
\botrule
\end{tabular} \label{ta3}}
\end{table}

The problem with our model (I) is the unusually large $\eta N$ branching ratio of 17\%
for the $D_{15}$ resonance, which strongly violates $SU(3)$, where an upper limit of
$2.5\%$ has been evaluated\cite{Polyakov:2006}. Furthermore, it may also be in conflict
with single and double pion photoproduction data on the neutron as well as with the
hadronic $\pi,\eta$ reactions, that are the more standard sources for branching ratios.
However, all those data are not good enough to draw definite conclusions. On the other
side, the reggeized model (II), which does not require a large $D_{15}$ coupling in order
to explain the cross section and beam asymmetry data cannot describe the bump structure
in the neutron data. This is drastically shown in the neutron to proton ratio in
Fig.~\ref{fig:CBeta} (last panel).
\begin{figure}[htp]
\centerline{
\includegraphics[width=10cm, angle=0]{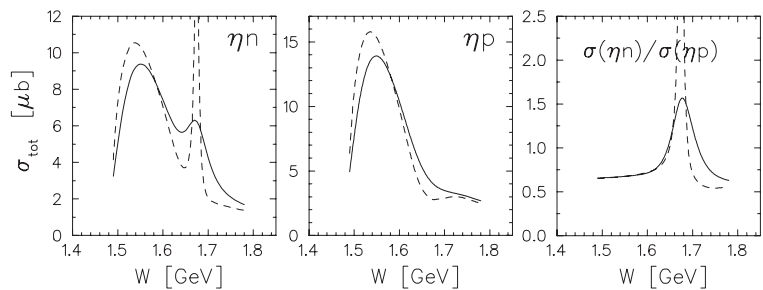} }
\caption{\label{fig:PQ} Eta photoproduction with a pentaquark
state with properties given in Table \ref{ta3}. The dashed lines show the production on a
free nucleon and the solid lines on a quasi-free nucleon of a deuteron target.}
\end{figure}
Alternatively, we have studied the excitation of a narrow $P_{11}$ pentaquark state in
the quasi-free eta photoproduction off the deuteron. Already Diakonov et
al.\cite{Diakonov:1997} in a chiral solition model and Arndt et al.\cite{Arndt:2004} in a
modified partial wave analysis had reported about a possible $P_{11}$ state in the region
around $1680-1730$~MeV with a width smaller than 30~MeV, that should couple quite
strongly to the $\eta N$ channel. Furthermore, Polyakov and Rathke\cite{Polyakov:2003}
have shown that the e.m. transition moment of the neutron should be much larger than for
the proton, with a ratio of $\mu_{n N^*}/\mu_{p N^*} \gtrsim 3$. Based on these
properties we have included this state as a $P_{11}(1675)$ resonance in our isobar model
with parameters given in Table~\ref{ta3} and have calculated the cross sections for both
a free proton and a free neutron target. The total cross sections can be seen in
Fig.~\ref{fig:PQ}, where for the neutron this state pops out of the background as a sharp
resonance, while it is hidden in the background for the proton target. The huge
difference can be understood from the fact, that the e.m. moments or couplings enter
quadratically in the cross sections, thus giving a ratio of one order of magnitude.

However, such a sharp resonance would not show up in an experiment on a bound neutron in
a deuteron target. Due to Fermi motion, the sharp resonance state becomes broadened and
gets a shape similar to an ordinary nucleon resonance with a width of around 100~MeV. The
solid line in Fig.~\ref{fig:PQ} shows a calculation in the spectator-nucleon
approach\cite{Fix:2006}. Further corrections of $NN$ and $\eta NN$ final state
interaction are expected to be small\cite{Fix:2006}.

\begin{figure}[ht]
\centerline{
\includegraphics[width=12.0cm, angle=0]{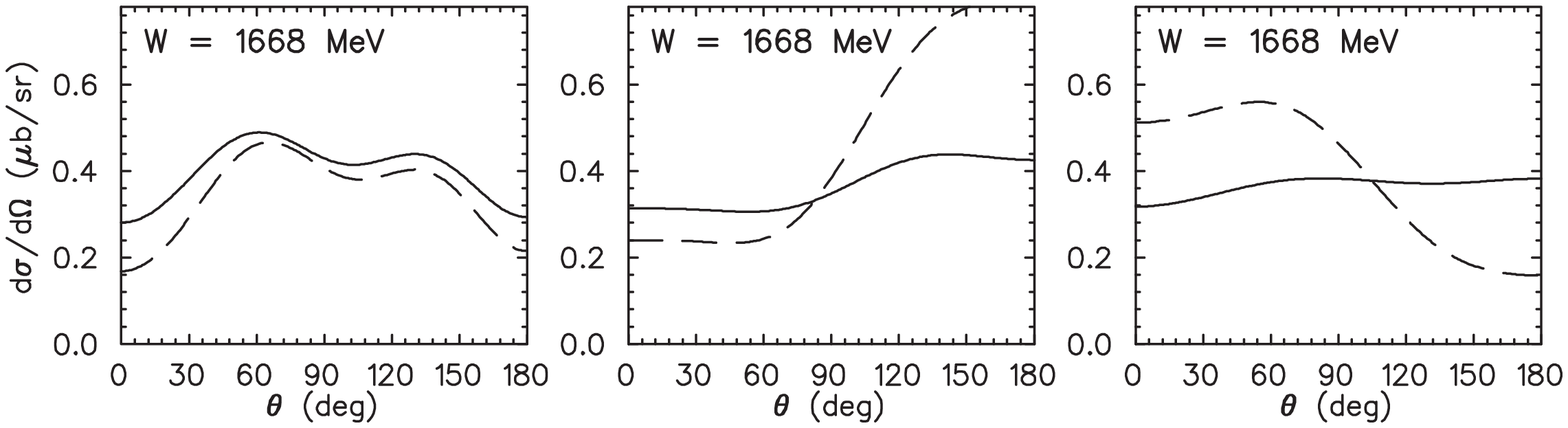} }
\caption{\label{fig:dsgpq} Differential cross section for eta photoproduction on the neutron
at $W=1668$~MeV. From left to right, the first panel shows the result of $\eta$-Maid2003
with standard vector meson poles and a strong $D_{15}(1675)$ resonance, the second panel
shows the result with the reggeized model and a narrow $P_{11}(1675)$ resonance with a
hadronic phase $\zeta_{\eta N}=+1$, the third panel is similar to the second but with a
negative hadronic phase. The dashed lines are for a free neutron target and the solid
lines for a quasi-free neutron of a deuteron target. The cross section on a quasi-free
neutron refers to the so-called effective $\gamma N^*$ system, where the initial nucleon
is assumed to be at rest in the deuteron. A more detailed description of this system can
be found in Ref.~\protect\refcite{Krusche:1995}.}
\end{figure}

While both reaction mechanisms give similar results for the total cross section, due to
the different orbital momentum of the $D_{15}$ and $P_{11}$ resonances, they will show up
with different angular distributions. In Fig.~\ref{fig:dsgpq} we show our calculations
for a neutron at $W=1668$~MeV for a) the strong $D_{15}$ model, b) the narrow $P_{11}$
model with a phase $\zeta_{\eta N}=+1$ and c) the narrow $P_{11}$ model with a phase
$\zeta_{\eta N}=-1$. For all three cases we compare the calculation for a
free neutron with the quasi-free calculation on the deuteron. In case a) the averaging
over the spectator nucleon due to Fermi motion gives only a small effect, whereas in the
cases b) and c) with the narrow pentaquark state the Fermi smearing is very large, in
particular at the chosen energy very close to the resonance peak. In the angular
distribution the hadronic phase becomes very important, since the $P_{11}$ partial wave
interferes with other partial waves like the $S_{11}$. In the total cross section it can
only interfere with other contributions in the same partial wave, e.g. from the
background, and the difference can hardly be seen.

\subsection{$\bm{\eta'}$ photoproduction results on protons}
\label{sec:eta'}%
The experimental data base for $\eta'$ photoproduction is still rather limited. Besides
the total cross section data measured decades ago at DESY, the only modern data were
obtained at SAPHIR-ELSA\cite{Plotzke:1998} and very recently at
JLab/CLAS\cite{Dugger:2006}. Further data for differential cross sections have been taken
at CB-ELSA which can be expected to come out soon.

The isobar model $\eta'$-MAID is conceptional very similar to the $\eta$-MAID.  The
vector meson couplings used in the $t$-channel exchanges are well determined: The photon
couplings can be obtained from the electromagnetic decay widths of $\eta' \rightarrow
\rho \gamma$ and $\eta' \rightarrow \omega \gamma$ and for the strong couplings the same
values as in $\eta$ photoproduction are used. Furthermore, we neglect the Born terms as
in the case of $\eta$ photoproduction. Therefore, the background contributions are
completely fixed and only the resonance parameters are varied to fit the data.

\begin{figure}[htp]
\centerline{
\includegraphics[width=6.3cm, angle=0]{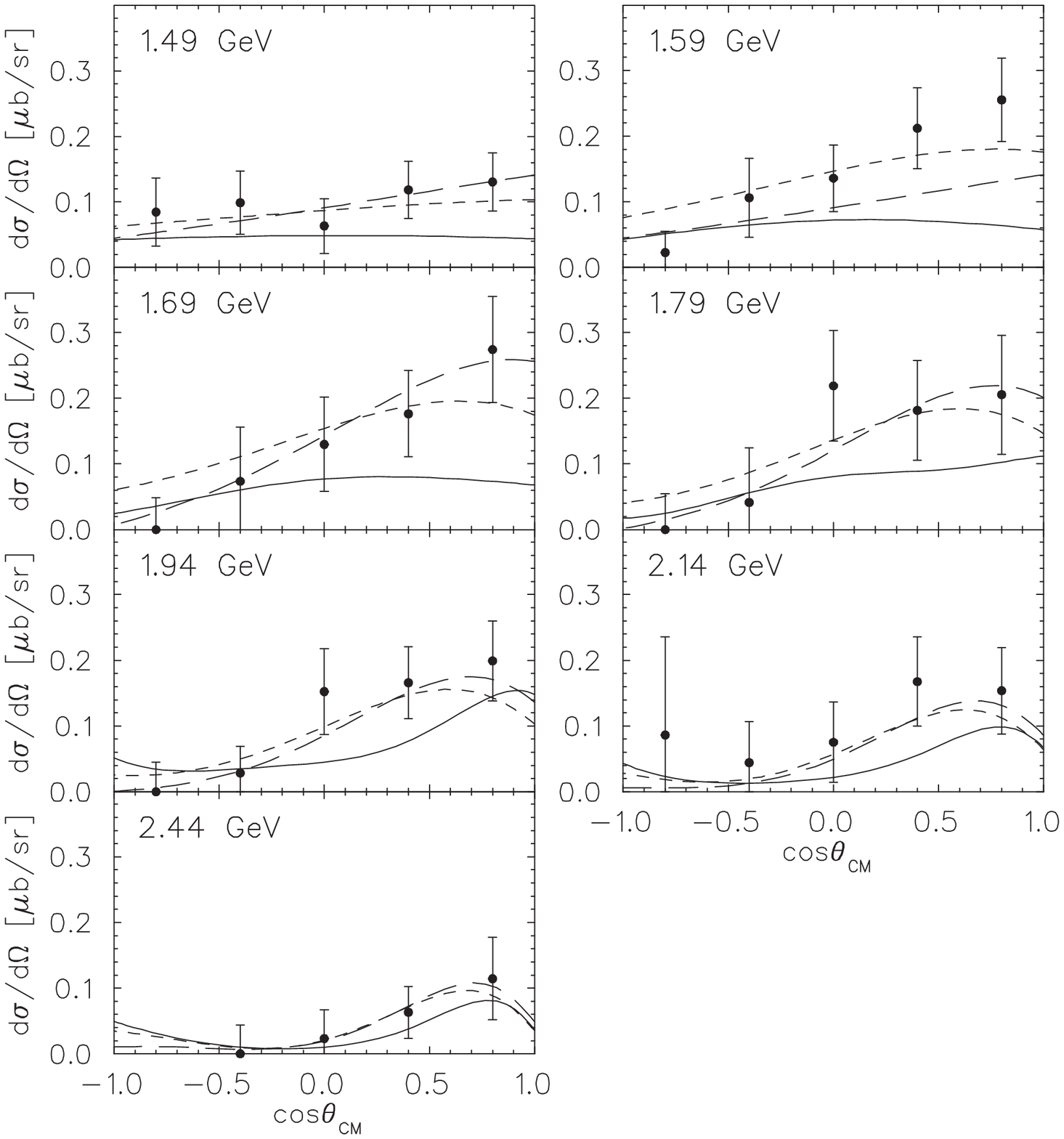}\hspace*{0.0cm}
\includegraphics[width=6.3cm, angle=0]{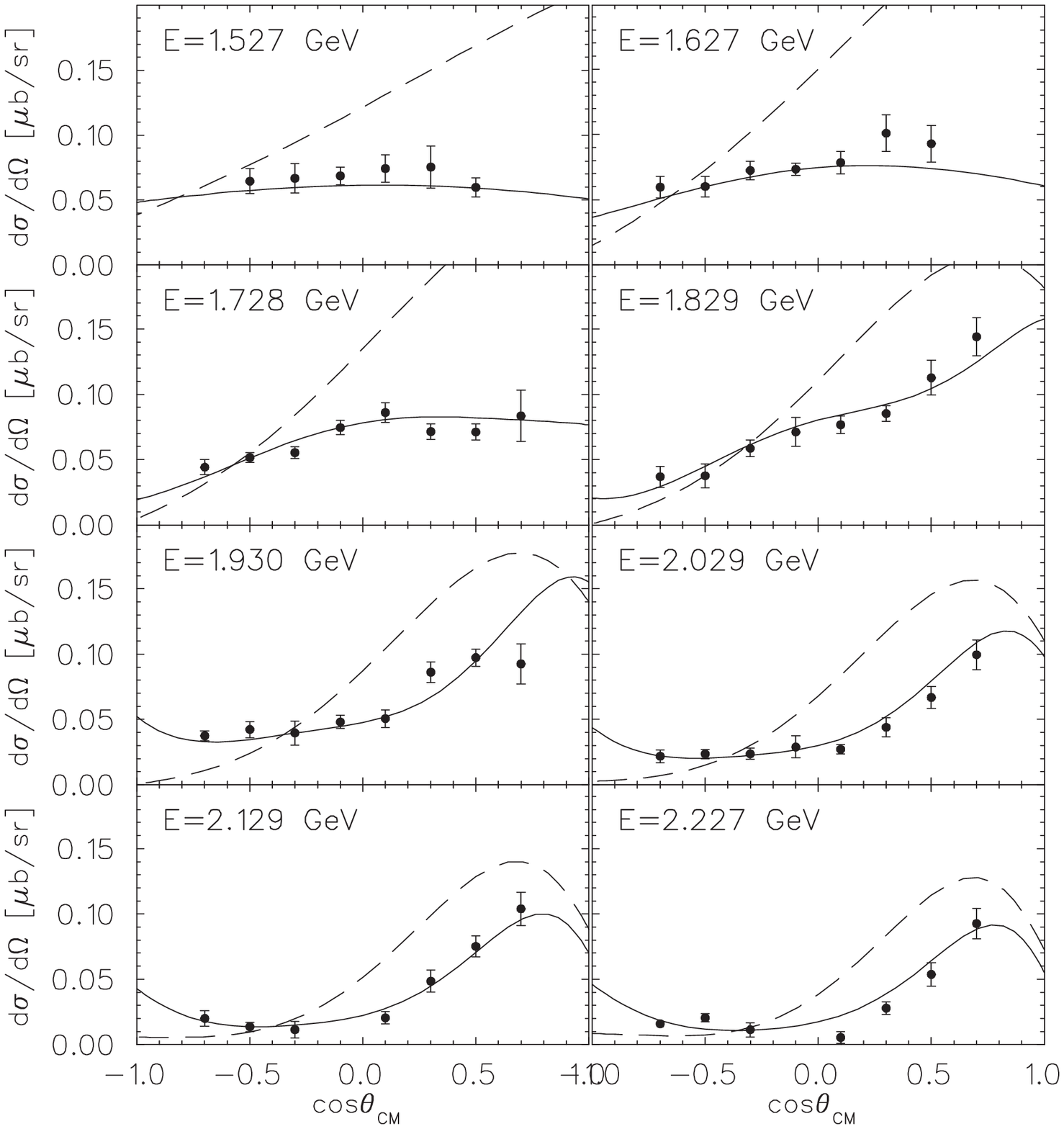}}
\caption{\label{fig:Etap} Etaprime photoproduction on the proton.
On the left we show the SAPHIR data of 1998 and on the right the CLAS data of 2006. The
dotted lines and the dashed lines show our one($S_{11}$)- and
two-resonance($S_{11},P_{11}$) fits from 2003 to the SAPHIR data only. The solid lines
are our new fits to the CLAS data including four resonances $S_{11},P_{13},P_{11},D_{13}$
as listed in Table~\ref{ta4}.}
\end{figure}

In our fit to the $\eta'$ data we use as free parameters the resonance mass and width as
well as the photon couplings $A_{1/2}$ and $A_{3/2}$. The also unknown branching ratios
into the $\eta' N$ channel cannot be fitted simultaneously. Therefore, we only determine
effective $p(\gamma,\eta')p$ couplings for both helicity states,
$\chi_{1/2}=\sqrt{\beta_{\eta'N}}A_{1/2}$ and $\chi_{3/2}=\sqrt{\beta_{\eta'N}}A_{3/2}$.
The results of our fit to the new CLAS data are given in Table~\ref{ta4}.
\begin{table}[ht]
\tbl{Mass, total width, and effective $p(\gamma,\eta')p$ couplings for total helicity 1/2
and 3/2 in units of $(10^{-3}/\sqrt{GeV})$ as defined in the text.}
{\begin{tabular}{ccccc} \toprule
resonance & $M^*$ (MeV)& $\Gamma_{tot}$(MeV) & $\chi_{1/2}$ & $\chi_{3/2}$\\
\colrule
$S_{11}$  & 1904 & 527 & 15.7 & -- \\
$P_{13}$  & 1926 & 146 & -1.5 & 1.0 \\
$P_{11}$  & 2083 & 51 & 2.5 & -- \\
$D_{13}$  & 2100 & 91 & 6.5 & -6.5\\
\botrule
\end{tabular} \label{ta4}}
\end{table}

From our previous fit to the SAPHIR data we concluded three solutions with a reggeized
vector meson background and a) an $S_{11}(1959)$ resonance, b) an $S_{11}(1932)$ and a
$P_{11}(1951)$ and c) an $S_{11}(1933)$ and a $P_{13}(1954)$. All of them give similar
$\chi^2$ with the old Bonn data but fail to describe the new CLAS data. Therefore, we
performed a new fit only to the CLAS data, shown in Fig.~\ref{fig:Etap} as the solid
line, that includes four nucleon resonances, $S_{11}, P_{13}, P_{11}$ and $D_{13}$. The
fitted parameters of these states are given in Table~\ref{ta4}. The $S_{11}$ and $P_{13}$
states cannot be found in the Particle Data Tables and could be identified with the
missing resonances that are claimed in many quark model calculations, e.g.
Refs.~\refcite{Capstick:1986,Giannini:2002,Metsch:2001}. The $P_{11}$ and $D_{13}$ states
around $W=2100$~MeV can be identified with the listed states $P_{11}(2100)$ and
$D_{13}(2080)$. This data is also well fitted within the relativistic meson-exchange
models of Sibirtsev\cite{Sibirtsev:2004} and Nakayama, Haberzettl\cite{Nakayama:2004},
including also the hadronic reaction $pp\rightarrow pp\eta'$. Besides the sub-threshold
resonances $S_{11}(1535)$, $P_{11}(1710)$, $D_{13}(1780)$ which contribute to the
background, the fit to the data also finds resonant contributions of $P_{13}(1940)$ and
$D_{13}(2090)$, see Ref.~\refcite{Dugger:2006}. Obviously, with only differential cross
section data many solutions with different resonances are possible and no definite
conclusions can be drawn at this stage.

\section{Summary and conclusions}
\label{sec:sum}%

In this paper we have presented a new partial wave analysis with recent data on $\eta$
and $\eta'$ photoproduction. The data give rise to interesting speculations about a
narrow $P_{11}$ resonance in $n(\gamma,\eta)n$ and missing resonances in
$p(\gamma,\eta')p$. However, these solutions are not unique. Further experimental
investigations are necessary in order to clarify the situations. Precise angular
distributions of quasi-free eta photoproduction on the deuteron could solve the question
about the pentaquark. In the case of etaprime production, polarization data, e.g. beam
asymmetry could be very helpful to better determine the partial wave contributions in
this reaction.

\section*{Acknowledgments}

I would like to thank Prof.~B.~Krusche and I.~Jaegle of the CB-ELSA collaboration and
Dr.~M.~Dugger from the JLab/CLAS collaboration for kindly sharing their data before
publication. In particular I am also grateful to Dr.~A.~Fix for his help in the
quasi-free eta photoproduction calculations on the deuteron. This work was supported by
the Deutsche Forschungsgemeinschaft (SFB 443).

%\section{References}

\end{document}